\documentclass[reprint,twocolumn,superscriptaddress, showpacs,floatfix]{revtex4-1}

\usepackage{epsfig}
\usepackage{amsmath}
\usepackage{exscale}
\usepackage{array}
\usepackage{units}
\usepackage{amssymb}
\usepackage{setspace}
\usepackage{bbm}
\usepackage{color}
\usepackage{algorithm}
\usepackage{algpseudocode}
\usepackage{color}

\begin{document}

\title{The description of strong correlation within self-consistent Green's function second-order  perturbation theory}

%\title{Magnetic exchange coupling parameters from noncollinear analytic derivatives: a novel method based on a perturbative approach to constrained SCF problems }

\author{Jordan J. Phillips\footnote{Corresponding Author:~philljj@umich.edu}}
%\email{blahblah}
\affiliation{Department of Chemistry, University of Michigan, Ann Arbor, Michigan 48109, USA}
%\author{Kermit the Frog}
%\email{blahblah}
%\affiliation{Department of Muppets, Sesame Street}

\author{Dominika Zgid}
%\email{phill3jj@cmich.edu}
\affiliation{Department of Chemistry, University of Michigan, Ann Arbor, Michigan 48109, USA}

\begin{abstract}
We report an implementation of self-consistent Green's function many-body theory within a second-order approximation (GF2) for application with molecular systems. This is done by iterative solution of the Dyson equation expressed in matrix form in an atomic orbital basis, where the Green's function and self-energy are built on the imaginary frequency and imaginary time domain respectively, and fast Fourier transform is used to efficiently transform these quantities as needed. 
We apply this method to several archetypical examples of  strong correlation, such as a H$_{32}$ finite lattice that displays a highly multireference electronic ground state even at equilibrium lattice spacing. In all cases GF2 gives  a physically meaningful description of the metal to insulator transition in these systems, without resorting to spin-symmetry breaking. Our results show that self-consistent Green's function many-body theory offers a viable route to describing strong correlations while remaining within a computationally tractable single-particle formalism.
\end{abstract}
\maketitle

\newpage

\section{Introduction}

The description of strong correlations arising from for example open shell $d$ orbitals in transition metal complexes\cite{B515732C,NewCorr4TransMetalChem2011}, or degenerate $\pi$ orbitals in polyaromatic organic compounds\cite{StrongCorrAceneSheetsJPCA2011}, remains a significant challenge in quantum chemistry for several reasons. For example, while single-reference methods such as Kohn-Sham density functional theory\cite{DFTbookParr} (DFT) can be cost-effective enough to apply to realistic systems, they flounder when confronted with multireference configurations that require a rigorous treatment of static correlation\cite{StaticCorrBurkejcp2005,ChallengesForDFT2012}. On the other hand, multireference methods that can handle these strong correlations, such as CASSCF \cite{CASSCF_1980,*CASSCF_1980_ii,*CASSCF_1981,CASPT2a,*CASPT2b} and RASSCF type methods\cite{RASSCF,RASPT2jcp2008}, or the more recently developed multiconfigurational hybrid DFT schemes\cite{Leininger:1997hs,longshortSavinjcp2007,MCDFTjcpSavin2012}, all share the same fundamental requirement of defining an active-space within which the number of determinants grows exponentially with the number of correlated orbitals.
 These challenges have motivated the exploration of alternate approaches for strongly correlated systems, such as the density matrix renormalization group (DMRG)\cite{TheDMRG,Chan_DMRG_jcp_2006,DMRGzgid2009,NewCorr4TransMetalChem2011},
 the two-electron reduced density matrix method\cite{2eRDMjcp2008,Hchainlattice2eRDMjcp2010,StrongCorrAceneSheetsJPCA2011}, constrained-pairing mean-field theory (CPMFT)\cite{CPMFT_jcp_2002,hlatticeCPmeanfieldScuseriajcp2009,CPMFT_II_2009} ,
  and projected Hartree-Fock theory (PHF)\cite{ProjectedHartreeFock2012,PHFplusDFT_2013jcp,PHF_jpca_2013,PHF_jpcb_2013,PHFjpca2014,PHF_Mo2_MolPhys_2014}. 
  Despite these developments, an inexpensive and generally applicable \emph{ab initio} method that can simultaneously handle dynamic and strong static correlation remains elusive. 
  %For example, DMRG, while performing excellently for low dimensional systems, faces tremendous obstacles in application to general three-dimensional materials. PHF offers a strong formalism to accurately treat static correlation at mean-field cost, but is not size-extensive and may not recover enough dynamic correlation for some applications. Similarly, CPMFT handles strong static correlation well but misses dynamic correlation.

%Methods based on the single-particle many-body Green's function, 
%\emph{e.g.} 
%dynamical mean-field theory\cite{DMFTfromQC2011zgid,GeorgesKotliarDMFT1996} and 

Green's function many-body theory\cite{fetter2003quantum} offers an interesting formalism to attack this problem.  In various realizations such as GW\cite{HedinGWpra1965}, the random phase approximation (RPA)\cite{RPA1,*RPA2,*RPA3,RPA_GellMann_Brueckner_1957},   the $n^{\textnormal{th}}$-order algebraic diagrammatic construction  (ADC{\footnotesize (n)})\cite{ADCn,*ADC_pra_1983,*ADC_pra_1989}, and second-order Green's function theory (GF2)\cite{mp2gf2HolleboomJCP1990}, many-body theory has a long history of use for calculating properties such as ionization potentials and electron affinities, excited states, spectra, and ground-state properties as well\cite{Doll_MB_GreensFunction_Finite_jcp_1972,OnGfAndTheirApplicationsCederbaum1990,Ortiz_Review_2013}.
%GW\cite{HedinGWpra1965} 
%RPA\cite{RPA1,*RPA2,*RPA3,RPA_GellMann_Brueckner_1957}
%ADC(n)\cite{ADCn}
%\cite{ADC_pra_1983}
%
While typically these methods have been applied to weakly correlated systems, it has  long been known that they can work with varying degrees of success for simple multireference systems. For example 
GW\cite{scGW4AtomsMOleculesEPL2006,VarEFunctionalGfPRA2006,levelsofscGWjcp2009Dahlen,scGWprb2013Scheffler}, RPA\cite{Furche_RPA_2001_prb,TotalEnergy_MBT_prl_2002,Commen_on_TotalEnergy_MBT_prl_2002,StaticCorrBurkejcp2005,RPA_Renorm_BatesFurche_2013,BondBreak_MBPT_prl_2013,Moussa_jcp_2014,StaticCorrelation_beyond_RPA_BSE_jcp_2014}, and GF2\cite{Dahlenjcp2005,mp2gf2HolleboomJCP1990} can  give a qualitatively correct description of stretched H$_{2}$  without breaking spin-symmetry.  Similarly, the ADC(n) method has been shown to accurately describe the spectra and excited states of multireference polyenes\cite{HowMuchDoublePolyene_ADC_2006} and carbon clusters\cite{CarbonCluster_ADC_jcp_1999}, and recently in the Nuclear Physics community second-order Gorkov-Green's function theory\cite{Gorkov,GF2_Soma_PRC_2011,*GF2_Soma_PRC_2014} has found use for open shell nuclei featuring degeneracies\cite{GF2_Soma_PRC_2013}.
%
%
%In this Communication it is our purpose to show that GF2 

In this  Communication it is our purpose to show that  GF2 can, without breaking spin-symmetry, give a qualitatively correct description of even
%include certain mulit-reference character and avoids the typical second order perturbation theory divergencies  for  
nontrivial strongly correlated systems such as stretched H$_{12}$ and H$_{32}$ lattices, which are multireference even at their respective equilibrium lattice spacings. We will show that in terms of calculated energies, GF2 is similar to MP2 (M\o ller--Plesset second order) when the system is single-reference, yet more closely resembles truncated CI in the strongly correlated dissociation limit. 
 Furthermore, at self-consistency GF2 yields fractional natural occupation numbers that can describe the metal-to-insulator transition in these lattices. 
This makes GF2 a viable formalism for systems too large for active-space methods, yet too strongly correlated for single-reference approaches, and offers a way to simply and efficiently extend MP2's  treatment of dynamic correlation to strongly correlated systems while remaining within a tractable single-reference formalism.

First we give an overview of the theory and our implementation of GF2. We stress that previous implementations of GF2-type methods by Holleboom and Snijders\cite{mp2gf2HolleboomJCP1990} as well as  Dahlen and van Leeuwen\cite{Dahlenjcp2005} 
%\cite{GF2_VanNeck_1991} \cite{GF2_Soma_PRC_2014} 
%\cite{Doll_MB_GreensFunction_Finite_jcp_1972}
existed, however to the best of our knowledge none of them investigated in detail  the potential of GF2 to describe highly mulitireference systems,  or an efficient implementation that can be useful by a more general community of computational chemists.
% with special attention to how our present implementation differs from that of Dahlen and van Leeuwen\cite{Dahlenjcp2005}.

\section{Implementation}

In the following we work in a  non-orthogonal atomic orbital (AO) basis with corresponding overlap matrix $\mathbf{S}$, Fock matrix $\mathbf{F}$, and density matrix $\mathbf{P}$.  Starting from a restricted Hartree-Fock (HF) reference solution,  in the AO basis the frequency dependent HF Green's function is built as $\mathbf{G}_{\textnormal{HF}}(\omega)=\bigr[(\mu+\omega)\mathbf{S}-\mathbf{F}\bigr]^{-1}$, where $\mu$ is the chemical potential, $\omega$ is an imaginary frequency, and $\mathbf{F}$ is given by

%\begin{equation}
%\mathbf{G}_{\textnormal{HF}}(\omega)=\bigr[(\mu+\omega)\mathbf{S}-\mathbf{F}\bigr]^{-1} ~,
%\label{eq:hfgf}
%\end{equation}

%\noindent where $\mu$ is the chemical potential, $\omega$ is a imaginary frequency, and $\mathbf{F}$ is given by

\begin{equation}
F_{ij}=h_{ij}+\underset{kl}{\sum}P_{kl}(\textnormal{v}_{ijlk}-\frac{1}{2}\textnormal{v}_{iklj})~,
\label{eq:FockMatrix}
\end{equation}
where $h_{ij}$ are  matrix elements of one electron operators and $\textnormal{v}_{ijkl}$ are two-electron integrals in the AO basis. %In our implementation, we choose to work on imaginary frequency axis described by the Matsubara grid $iw_n=i\frac{(2n+1)\pi}{\beta}$, where $\beta$ is the inverse temperature. 

\begin{figure}
\includegraphics[width=8.5cm]{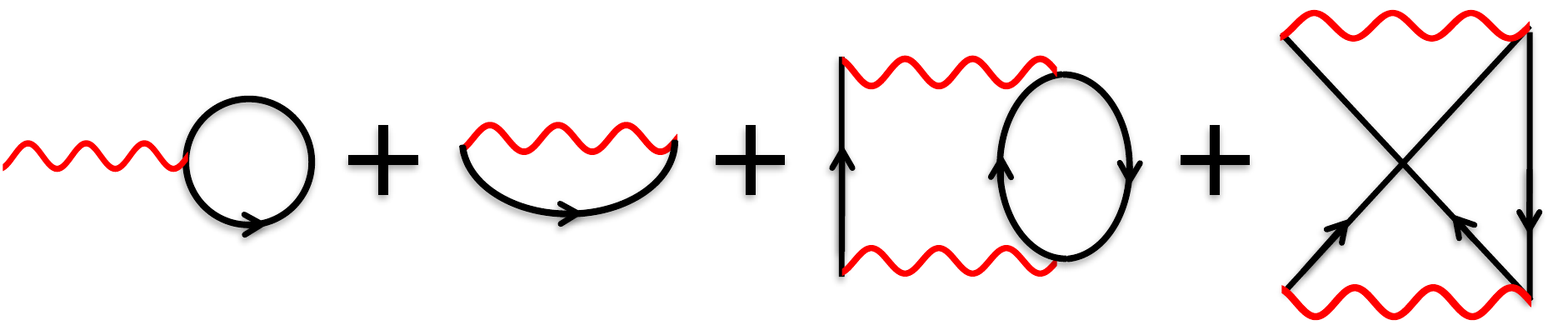}
\caption{Diagrams included in second-order self-energy within GF2. The first two diagrams are frequency-independent, and are included in the Fock matrix. The next two diagrams are frequency-dependent, and represent the second-order correlation effects covered by $\mathbf{\Sigma}(\omega)$.}
\label{fig:diagrams}
\end{figure}

 Provided with $\mathbf{G}_{\textnormal{HF}}$, the exact single-particle many-body Green's function, $\mathbf{G}(\omega)$, can be found by solving the Dyson equation $\mathbf{G}(\omega)=\mathbf{G}_{\textnormal{HF}}(\omega)+\mathbf{G}_{\textnormal{HF}}(\omega)\mathbf{\Sigma}(\omega) \mathbf{G}(\omega) $,
%
%\begin{equation}
%\mathbf{G}(\omega)=\mathbf{G}_{\textnormal{HF}}(\omega)+\mathbf{G}_{\textnormal{HF}}(\omega)\mathbf{\Sigma}(\omega) \mathbf{G}(\omega) ~,
%\label{eq:dyson}
%\end{equation}
%
where $\mathbf{\Sigma}(\omega)$ is the exact frequency-dependent self-energy which  accounts for the correlation missing in the simple HF picture. 
%Of course, an exact treatment of $\mathbf{\Sigma}(\omega)$ is not feasible and for practical calculations one must make approximations. 
%When the self-energy is approximated as a functional of the Green's function, $\mathbf{\Sigma}[\mathbf{G}]$, this results in a self-consistent scheme for solving Eq.~\ref{eq:dyson}. 
In this work we employ a second-order approximation to the self-energy, which is shown using Feynman diagrams\cite{Mattuck_Feynman} in Figure~\ref{fig:diagrams}.  Starting from the left the first two  are the first-order Hartree and exchange diagrams. %included in the $\Sigma_{\infty}$ 
This is the frequency independent part of the self-energy, and is already covered by $\mathbf{F}$ in Eq.~\ref{eq:FockMatrix}. The next two diagrams are second-order, and are given algebraically in the time-domain as

\begin{equation}
\begin{split}
\Sigma_{ij}(\tau)=-\underset{klmnpq}{\sum}G_{kl}(\tau)G_{mn}(\tau)G_{pq}(-\tau)\\
\times\textnormal{v}_{imqk}\bigr(2\textnormal{v}_{lpnj}-\textnormal{v}_{nplj}\bigr)~,
\end{split}
\label{eq:Sigma}
\end{equation}
%\noindent In Eqs.~\ref{eq:FockMatrix} and \ref{eq:Sigma} $h_{ij}$ are  matrix elements of one electron operators,  $\textnormal{w}_{ijkl}$ are two electron integrals in the AO basis,
%\cite{Szabo:1989rc}
 where $\mathbf{G}(\tau)$ is the Green's function Fourier transformed to the imaginary time domain.  With the self-energy constructed, we can fast Fourier transform (FFT)  $\mathbf{\Sigma}(\tau)$ to the $\omega$ domain and build the Green's function as

\begin{equation}
\mathbf{G}(\omega)=\bigr[(\mu+\omega)\mathbf{S}-\mathbf{F}-\mathbf{\Sigma}(\omega)\bigr]^{-1}~.
\label{eq:GF}
\end{equation}

\noindent Provided with an updated Green's function, we can build the correlated single-particle density matrix, $\mathbf{P}=\frac{1}{2\pi i}\oint \mathbf{G}(\omega)d\omega$,
 and then update the Fock matrix according to Eq.~\ref{eq:FockMatrix}.
%Given some $\mathbf{G}(\omega)$, the Green's function determines the density matrix $\mathbf{P}$ and therefore the Fock matrix by Eq.~\ref{eq:FockMatrix}.  In addition, after fast Fourier transform (FFT) to the $\tau$ domain, $\mathbf{G}(\tau)$ determines the self-energy $\mathbf{\Sigma}(\tau)$ according to Eq.~\ref{eq:Sigma}. Once built, $\mathbf{\Sigma}(\tau)$ can be Fourier transformed back to the frequency domain, yielding $\mathbf{\Sigma}(\omega)$.  
Equations \ref{eq:FockMatrix}, \ref{eq:Sigma}, and \ref{eq:GF} therefore establish a self-consistent scheme for calculating the Green's function in a second-order approximation. 
A cartoon showing a bird's-eye view of our implementation of the GF2 algorithm is presented in Figure~\ref{fig:loops}.  The detailed description of the implementation steps can be found in the Supplementary Material\cite{supmat_GF2Paper}.

Let us now discuss several advantages of the GF2 theory that may make it accessible and interesting for a general quantum chemistry community. 
GF2 is an iterative procedure 
where the central quantity of interest is the single-particle many body Green's function $\mathbf{G}(\omega)$, rather than the single-particle density $\rho (\mathbf{r})$ or the many-body wavefunction $\Psi$. 
% since one can treat different grid (frequency) points in a massively parallel fashion on different processors.
In the GF2 scheme there is no explicit reference to  $\Psi$,  thus  the requirement of choosing an active-space or truncating $\Psi$ at a certain excitation level is circumvented entirely. This offers an enormous reduction in computational effort for systems that require large active-spaces, and gives GF2 generous flexibility in handling systems that would require different levels of excitations  in a CI expansion (\emph{e.g.} singles, doubles, triples, \emph{etc.}). Additionally, unlike KS-DFT, which depends explicitly on a $\rho (\mathbf{r})$ built from a non-interacting reference of Kohn-Sham orbitals obeying Aufbau filling, GF2 has no problem confronting strongly correlated systems and can yield fractional natural occupation numbers at self-consistency.
Furthermore, because of its iterative self-consistent nature GF2 is independent of the reference Green's function, and thus both HF and DFT starting Green's functions can be used. 
  In the final solution, series of diagrams are included due to the inexplicit resummation in the iterative procedure. This is the reason why GF2 is able to recover some static correlation  even when starting from a restricted-HF solution, and can avoid the typical MP2 divergence for cases with decreasing band gaps.   
  %This diagrammatic resummation that automatically occurs during the GF2 iterations has further dramatic consequences and 
 Moreover, due to the inclusion of series of mosaic diagrams this method is applicable to metallic systems\cite{footnote,CC_channels_egas_2014_scuseria,*RangeSeparated_CCSD_scuseria_2014} when the regular MP2 method would remain pathologically divergent.  

The building of $\mathbf{\Sigma}(\tau)$ according to Eq.~\ref{eq:Sigma} formally scales as $O(N_{\tau}n^{5})$, where $N_{\tau}$ and $n$ are the number of imaginary time grid points and atomic orbitals respectively. 
%
%$N_{\tau}$ presents a significant prefactor, and while in the past when serial computation was the norm rendered self-consistent GF2 calculations exceedingly expensive, in the present day of parallel computation this obstacle is essentially removed.
%The Green's function methods have not become mainstream in the quantum chemistry community since the price paid for this is contending with a frequency (or time) grid to numerically build the Green's function and self-energy. 
%While in the past this presented a serious obstacle when serial computation was the norm, in the present day of parallel computation this obstacle is essentially removed.
%
While $N_{\tau}$ presents a significant prefactor, at any given grid point $\tau_{n}$ the construction of the self-energy $\mathbf{\Sigma}_{\tau_{n}}$ is independent, and therefore as a whole the self-energy formation is embarrassingly parallel.  Given the development in technology in the last few decades, it is not unreasonable to have access to a computing facility offering several hundred to tens of thousands of processors for parallel computing. Therefore, practically speaking, the formal scaling of $O(N_{\tau}n^{5})$ will never be realized in typical calculations, and the reality will be much closer to $O(n^{5})$. 

Since the self-energy in Eq.~\ref{eq:Sigma} can be expressed in the imaginary time domain as a product of the Green's functions with the two-electron integrals, the whole GF2 calculation can be performed in the AO basis. Thus the costly orbital transformations from AO to molecular orbital (MO) basis are not necessary, and one can take advantage of integral screening which should reduce the computation cost even further. This is due to a similar structure of the second order self-energy in the time domain to the Laplace transformed MP2 expressions. 
%
% Furthermore, because our implementation of GF2 is  in an AO basis from the beginning, one can take advantage of integral screening which should reduce the computation cost even further. 
 An alternative route leading to a significant cost reduction is using the density fitted AO integrals in the Eq.~\ref{eq:Sigma} that would result in $O(n^{3}m)$ scaling where the $m$ is the number of auxiliary functions necessary for density fitting. Thus, as expected the overall scaling of GF2 algorithm is identical up to a prefactor with the Laplace transformed density fitted MP2 scheme. 

To build $\mathbf{\Sigma}(\tau)$ we  employ a non-equidistant $\tau$ grid, and we find that around 3000--4000 grid points is sufficient for the Fourier transforms to be converged to a very high precision\cite{ALPS}. We use  fast Fourier transforms that scale as $O(Nlog(N))$, where $N$ is the number of grid-points.  High accuracy in the FFT integrals is maintained  since we use an analytical high frequency tail of the Green's function\cite{Armin,Gull}, given by $G(\omega)\approx \frac{G_1}{\omega}+\frac{G_2}{\omega^2}+\frac{G_3}{\omega^3}+\dots$
%\begin{equation}
%G(\omega)\approx \frac{G_1}{\omega}+\frac{G_2}{\omega^2}+\frac{G_3}{\omega^3}+\dots
%\end{equation}
 %{\color{red}(we need to define G1,G2,G3?)}
To build $\mathbf{G}(\omega)$ we use a grid of Matsubara frequencies, $\omega_{n}=(2n+1)i\pi /\beta$. We find it useful to choose the inverse temperature $\beta$ and number of frequency points such that the initial Hartree-Fock Green's function reproduces the Hartree-Fock energy and particle number to a reasonable precision (the calculations reported here use $\beta=200-300$). This brings us to the next advantage of the GF2 method, which is the inverse temperature $\beta$ makes the GF2 calculation explicitly temperature dependent. We use such a high value of the inverse temperature since we are approximating regular quantum chemical calculations at $0$ K. However, if we were interested in calculating quantities at room temperature then our inverse temperature will have much lower values and the number of required both frequency and tau points would become much smaller.  
%Consequently, we believe that GF2 would be immensely useful for studying systems where temperature plays a non-trivial role. 
All other computationally demanding steps of the GF2 algorithm such as constructing of the Green's function and evaluation of the chemical potential scale as $O(n^3)$ and are cheaper than the construction of $\mathbf{\Sigma}(\tau)$. 
\begin{figure}
\includegraphics[width=4.5cm]{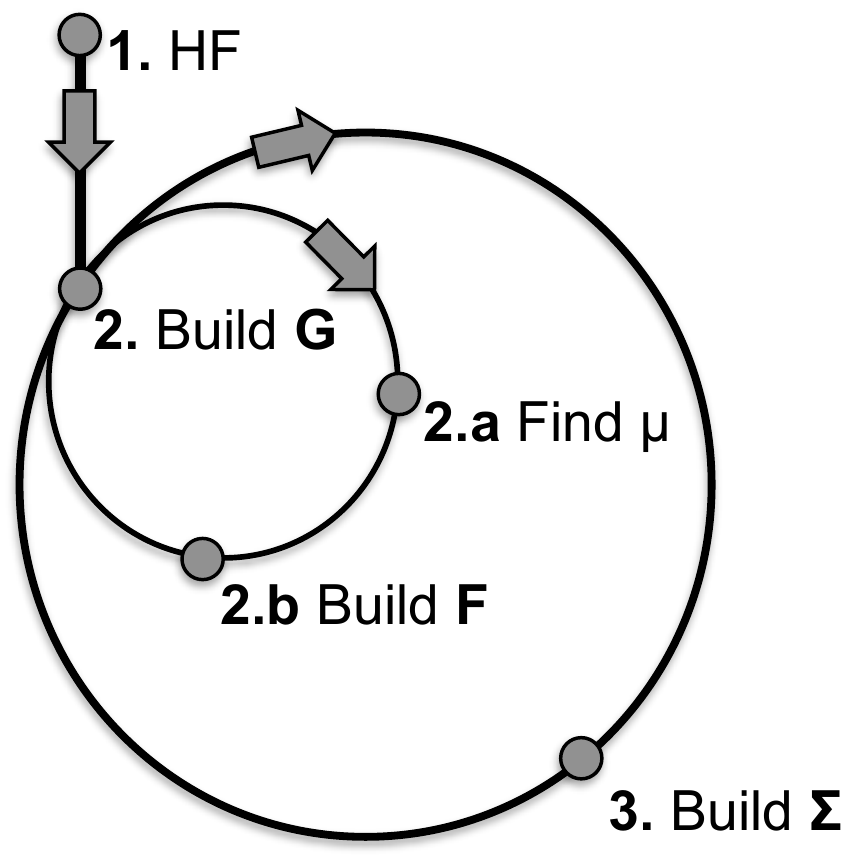}
\caption{A bird's-eye view of the GF2 algorithm.}
\label{fig:loops}
\end{figure}

%Finally, because of the self-consistent nature of GF2  it is not strictly necessary to begin from a Hartree-Fock reference, as in-principle the self-consistent Green's function should not depend on the initial Green's function\cite{Dahlenjcp2005}. Hypothetically, there could arise cases where $\mathbf{G}_{HF}$ is a poor starting reference. For example with trapped mixed-valence dinuclear Mn  complexes Hartree Fock has been shown to give a qualitatively wrong description of the $d$ orbital occupations, whereas hybrid DFT gave a qualitatively correct description\cite{PhillipsSemilocal2012}. In cases like this one could trivially begin the GF2 algorithm with DFT input, though it would need to be done carefully to avoid double-counting of correlation. For example, if one naively built $\mathbf{G}$ using the Kohn-Sham Fockian $\mathbf{F}_{KS}$, and then built the self-energy according to Eq.~\ref{eq:Sigma}, there would be a double-counting of correlation between $\mathbf{F}_{KS}$ and $\mathbf{\Sigma}$. This double-counting could be simply avoided by at first iteration using the converged Kohn-Sham density matrix $\mathbf{P}_{KS}$ to build $\mathbf{F}(\mathbf{P}_{KS})$ according to Eq.~\ref{eq:FockMatrix}, and then proceeding as usual with the GF2 algorithm.

%\section{Computational Details}

%We use Dalton to do these calculations\cite{Dalton}(I have to fix special characters in author list). 
%Cite ALPs? Cite STAMPEDE?

We emphasize  that this GF2 procedure is all-electron, with no selection of an active space of correlated orbitals. Other than the choice of an appropriate frequency/time grid (which usually is straightforward), the GF2 procedure requires no more user input than that of a typical HF or DFT calculation. As such it can be made blackbox. %Furthermore, the self-energy in GF2 is explicitly defined by strict diagrammatic methods, and is therefore expressly \emph{ab initio}. 
%Thus, one of the most obvious uses of GF2 procedure may be to use it to determine active space orbitals for complicated strongly correlated molecules studied with density matrix renormalization group (DMFT) or some of the QM/QM embedding methods such as dynamical mean field theory (DMFT). Another possibility is to use it in embedding scheme in order to choose the strongly correlated orbitals and treat the weakly correlated ones at the GF2 level.  Below, we describe our finding describing how well the GF2 method recovers both dynamic and static correlation avoiding the typical MP2 collapse for the strongly correlated cases.  
%\cite{GeorgesKotliarDMFT1996,DMFTfromQC2011zgid}

\section{Results and Discussion}

Now we report the application of our  GF2 implementation for several representative cases of strong correlation, and compare our results to standard \emph{ab initio} methods (\emph{e.g.} RHF, MP2, truncated CI, CCSD, \emph{etc.}). We want to stress that all GF2 calculations are spin-restricted and are compared to other spin-restricted calculations using standard \emph{ab initio} methods. The  DALTON 2011 suite of programs\cite{Dalton} was used to carry out all calculations, as well as to generate the restricted Hartree-Fock input required for our in-house GF2 implementation.

Before we discuss the dissociation curves, it is worth making a very brief remark on the relevance of self-interaction error (SIE), which can manifest in many-body theory \cite{Casida_pra_1991,SIE_GW_2007_pra,SIE_and_StaticCorrelation_Henderson_2010} and DFT\cite{SIEcPerdewZunger1981}. Because SIE can ``mimic'' static correlation\cite{PoloSIEcorrelation2002,SIE_and_StaticCorrelation_Henderson_2010}, its presence (absence) can actually appreciably improve (worsen) the performance of a method in the dissociation limit\cite{RPA_Renorm_BatesFurche_2013,BondBreak_MBPT_prl_2013}. Because GF2 includes all direct and exchange diagrams up to second order it is therefore explicitly SIE-free, in contrast to other many-body methods such as GW or RPA\cite{SIE_GW_2007_pra,SIE_and_StaticCorrelation_Henderson_2010}. Therefore the static correlation GF2 recovers  in our calculations is the result of a genuine treatment of correlation rather than a fortuitous exploitation of SIE errors. 
%Because SIE is well-known to be deleterious to DFT's applicability to molecular systems\cite{OxidationEnergies_SIE_PRB_2006}, it is not unreasonable to expect that being free of SIE is important in many-body theory as well.

\begin{figure}
\includegraphics[width=8.5cm]{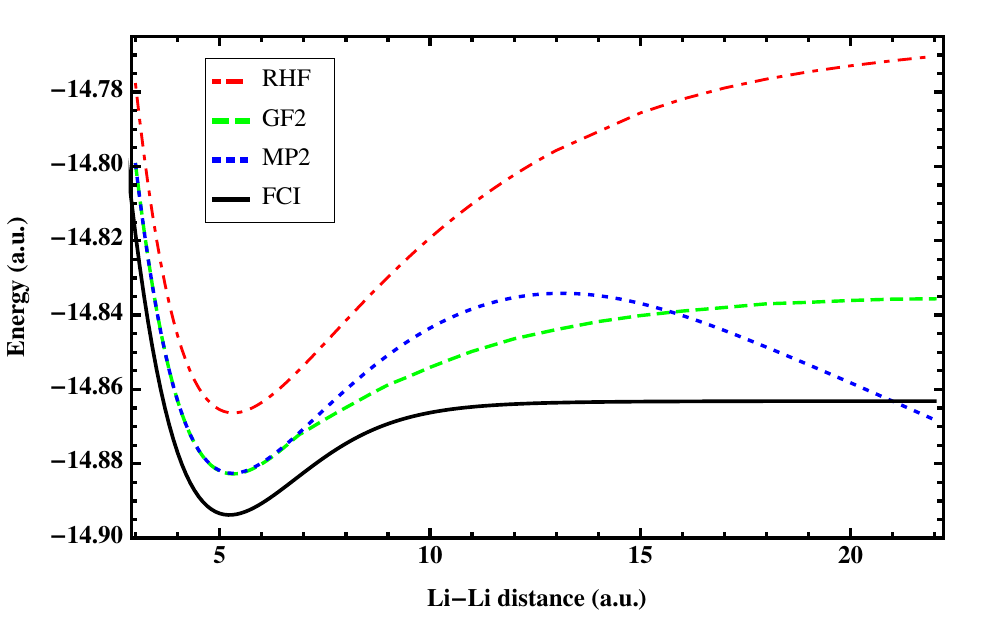}
\caption{Dissociation of Li$_{2}$ with 6-31G basis.}
\label{fig:LiHstretch}
\end{figure}

First we consider the Li$_{2}$ dissociation with the 6-31G basis, shown in Figure~\ref{fig:LiHstretch}. 
%Our GF2 calculations use $\beta =300$, with a Matsubara  $\omega$ grid of 12,000 points, and a non-equidistant $\tau$ grid of 4,400 points. 
For this system we compare RHF, MP2, and FCI to GF2. The natural occupation numbers obtained with GF2 for this system (as well as the other systems studied in this work) can be found in the Supplementary Material\cite{supmat_GF2Paper}. Interestingly,  MP2 and GF2 are practically identical within the equilibrium distance and up to about 7 a.u.
%, beyond which they gradually separate as the GF2 natural occupation numbers begin to show the onset of strong correlation. 
Beyond 7 a.u. the MP2 and GF2 curves separate, and the GF2 solution transitions from weakly to strongly correlated. MP2, unable to cope with the multireference character of the strongly correlated solution, characteristically begins to diverge to $-\infty$ correlation energy with infinite separation while GF2 converges to a finite value parallel to FCI. 

\begin{figure}
\includegraphics[width=8.5cm]{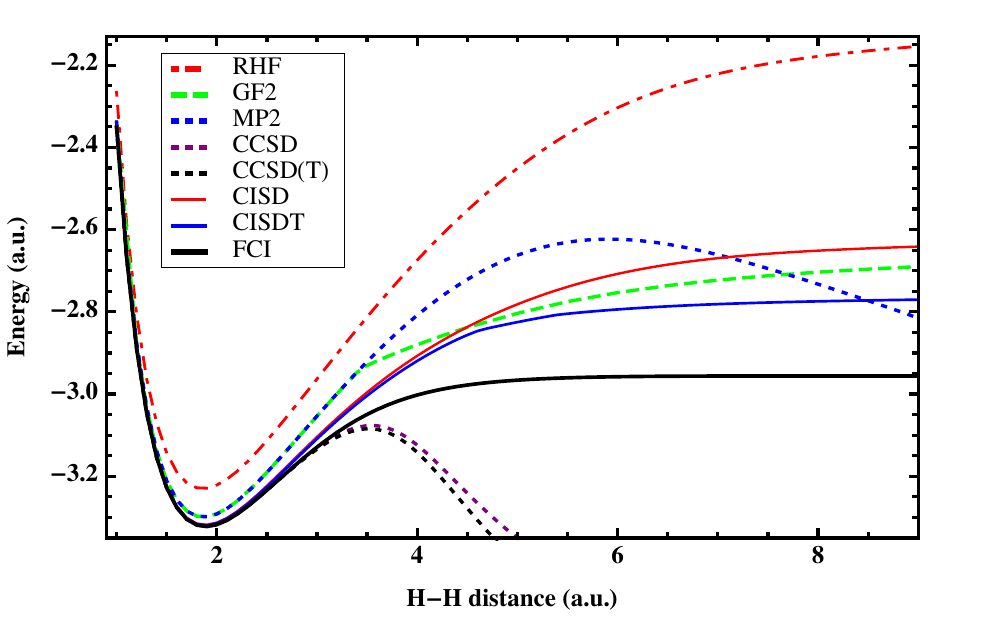}
\caption{Dissociation of H$_{6}$ with TZ basis.}
\label{fig:h6stretch}
\end{figure}

Next, we consider the H$_{6}$ ring stretch using a TZ basis, which is shown in Figure~\ref{fig:h6stretch}. Due to the high multireference character in the dissociation limit  both  CCSD and CCSD(T) fail spectacularly.  MP2 and GF2 are in very close agreement around the equilibrium distance where the system is still essentially single-reference. Beyond $a=3.5$~a.u. the system rapidly takes on multi-reference character and the GF2 curve  breaks away from MP2, which coincides with the divergence of CCSD and CCSD(T) away from the FCI curve. In the dissociation limit GF2 settles out between the CISD and CISDT curves. 

Now we apply GF2 to two finite hydrogen lattices, a 4$\times$3 hydrogen plaquette  (H$_{12}$) and a 4$\times$4$\times$2 hydrogen lattice (H$_{32}$).
 %which is bi-layer 4$\times$4 hydrogen plaquette (H$_{32}$). 
 The structures of both systems are determined by the lattice parameter $a$. Hydrogen chains and lattices are an interesting test case because they can display highly multireference electronic ground states at stretched lattice parameters, and accounting for the metal-to-insulator transition with changing lattice parameter is a challenging test for an \emph{ab-initio} method\cite{StaroverovScuseriaHFB2002jcp,Chan_DMRG_jcp_2006,Hchainlattice2eRDMjcp2010,StrongCorrelation_MonteCarlo_prb_2011}.

First we consider the 4$\times$3 hydrogen plaquette with a minimal STO-3G basis. This system is interesting as it displays strong correlations even at the equilibrium lattice spacing. Because of this MP2 and GF2 begin to separate from each other near the bottom of the well around 2.3 a.u., while CCSD and CCSD(T) cannot even qualitatively describe the equilibrium energy curve. In this case in the dissociation limit GF2 gives an energy very similar to CISDTQ.  Excepting of course FCI, of all the methods considered \emph{only} GF2 was able to capture the physically correct limit with all 12 natural occupation numbers approaching unity in the dissociation limit.

\begin{figure}
\includegraphics[width=8.5cm]{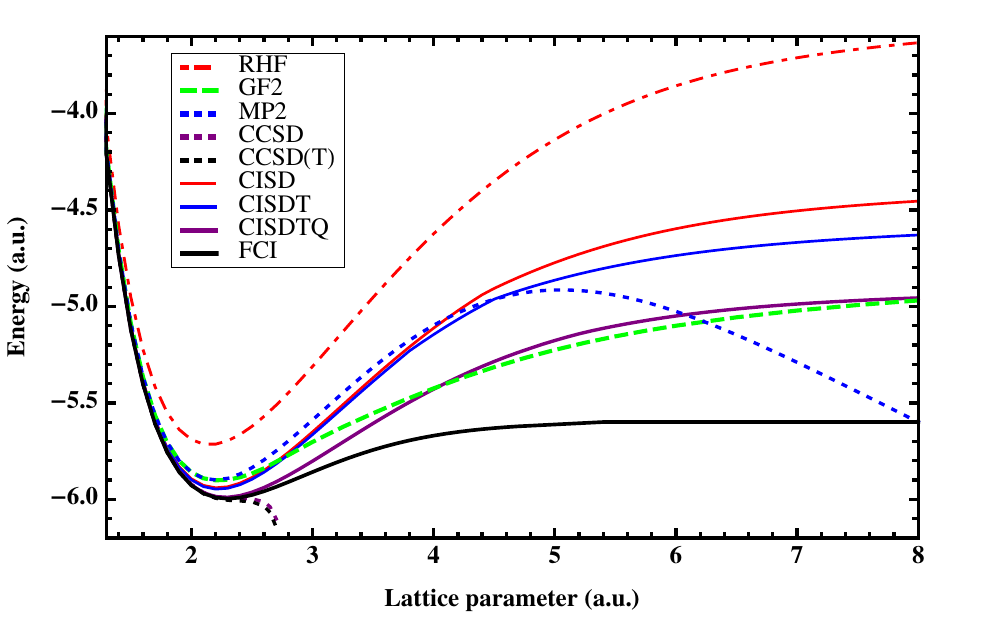}
\caption{Dissociation of H$_{12}$ with STO-3G basis.}
\label{fig:H12stretch}
\end{figure}

\begin{figure}
\includegraphics[width=8.5cm]{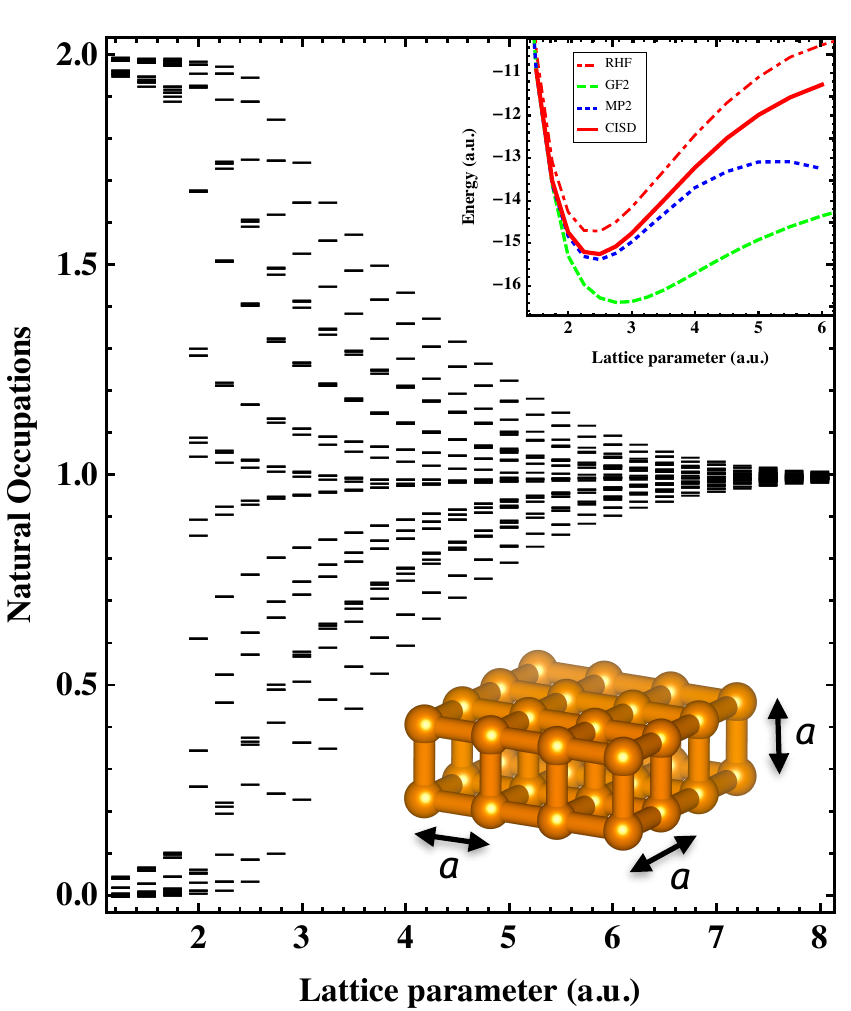}
\caption{Natural occupations numbers of H$_{32}$ double plaquette with respect to lattice parameter $a$. Inset lower-right: structure of H$_{32}$ (image generated by VESTA\cite{VESTA}). Inset upper-right: lattice stretching curves for RHF (dot-dashed red), MP2 (dotted blue), CISD (thick red), and GF2 (dashed green).}
\label{fig:h32}
\end{figure}

Finally, we consider the H$_{32}$ finite cubic lattice, which is even more strongly correlated within the equilibrium lattice spacing than H$_{12}$. In Figure~\ref{fig:h32} we plot the natural occupation numbers of the correlated density-matrix as-obtained by GF2 with respect to the lattice parameter $a$. Additionally, in the lower-right inset we show the lattice structure, and in the upper-right inset we show dissociation curves for RHF, MP2 , CISD, and GF2. We attempted coupled cluster calculations on this system but could not converge for $a\ge$~2.0~a.u., similar to Hachmann, Cardoen, and Chan's experience with the H$_{50}$ chain\cite{Chan_DMRG_jcp_2006}.
Examining first the natural occupation numbers,  for  $a<$~2.0~a.u. the system is approximately single-reference, but it's visible the GF2 natural occupations are smoothly shifting towards stronger correlations. By $a\approx$~2.0~a.u. a weakly to strongly correlated transition has occurred, and for
 $a\ge$~2.0~a.u. the system smoothly becomes more strongly correlated, with all 32 natural occupation numbers approaching unity in the limit of large lattice parameter $a$. This shows that GF2 can capture the metal-to-insulator transition in a strongly-correlated system while being essentially a single-reference-type blackbox method. 
 Inspecting the lattice-stretching curve in the upper-right inset now, we see that already 
 by $a=$~2.0~a.u. the MP2 and GF2 curves have separated, because of the strong correlations present near the equilibrium lattice spacing. Interestingly, the CISD curve is slightly worse than MP2, which indicates the truncated CI expansion of $\Psi$ would need to include much higher excitations than merely doubles, which is consistent with the natural occupation numbers from GF2.

%It should be noted that a system requiring an active space of 32 electrons in 32 orbitals is not merely an academic exercise. One of us recently studied a molecular magnet that would require a minimal active space of 35 $d$ electrons\cite{Fe7disks}. Strongly correlated systems with large active spaces will likely continue to be of scientific and technological interest in a broad range of fields.

In conclusion,  GF2  is an all-electron, size-extensive, \emph{ab initio} blackbox method that due to iterative diagrammatic resummation can recover both dynamic and static correlation. This allows it to avoid the typical collapse  inherent to the MP2 method when strong correlations are present, and furthermore enables GF2 to describe phenomena such as the metal-to-insulator transition in strongly correlated electronic systems. Additionally, GF2 gives immediate access to temperature dependent properties, and is a good formalism for calculating frequency dependent quantities as well. Moreover,  GF2 can be trivially implemented in a massively parallel fashion, and it can scale as $O(n^3m)$ with density fitted integrals, thus making it affordable. Further work is necessary to provide additional validation of GF2's performance for more realistic systems, but even in its current form this method can be used to help determine the active space orbital choice or provide a starting point for procedures such as dynamical mean field theory (DMFT)\cite{GeorgesKotliarDMFT1996,DMFTfromQC2011zgid}. 
Due to its low computational scaling, blackbox nature, ability to recover  multireference character and AO-based implementation, GF2 also makes an ideal candidate for extension to periodic systems. One can also dare to conjecture, based on some earlier results  with selective CC resummation for the electron gas\cite{CC_channels_egas_2014_scuseria,*RangeSeparated_CCSD_scuseria_2014}, that a periodic GF2 implementation would be applicable to both metals and insulators.

\section{Acknowledgments }

D. Zgid and J.J. Phillips acknowledge support from a DOE grant no. ER16391 and an XSEDE allocation allowing us to use the STAMPEDE supercomputer.

\section{Supplementary material for the manuscript \\ ``The description of strong correlation within self-consistent Green's function second-order  perturbation theory''}
\subsection{Implementation of GF2 algorithm}

The GF2 algorithm is defined by the following three self-consistent equations:

\begin{equation}
\mathbf{G}(\omega)=\bigr[(\mu+\omega)\mathbf{S}-\mathbf{F}-\mathbf{\Sigma}(\omega)\bigr]^{-1}
\label{eq:GF}
\end{equation}

\begin{equation}
F_{ij}=h_{ij}+\underset{kl}{\sum}P_{kl}(\textnormal{v}_{ijlk}-\frac{1}{2}\textnormal{v}_{iklj})
\label{eq:FockMatrix}
\end{equation}

\begin{equation}
\begin{split}
\Sigma_{ij}(\tau)=-\underset{klmnpq}{\sum}G_{kl}(\tau)G_{mn}(\tau)G_{pq}(-\tau)\\
\times\textnormal{v}_{imqk}\bigr(2\textnormal{v}_{lpnj}-\textnormal{v}_{nplj}\bigr)
\end{split}
\label{eq:Sigma}
\end{equation}

\noindent The iterative solution of the Dyson's equation according to Equations \ref{eq:FockMatrix}, \ref{eq:Sigma}, and \ref{eq:GF} proceeds as follows:

\begin{itemize}
\item[\textbf{1.}] Perform a restricted Hartree-Fock calculation, obtaining the HF Fock matrix $\mathbf{F}_{\textnormal{HF}}$, density matrix $\mathbf{P}_{\textnormal{HF}}$, and overlap matrix $\mathbf{S}$.
\item[\textbf{2.}] Build $\mathbf{G}(\omega)$ according to Eq.~\ref{eq:GF}. Assuming a Hartree-Fock reference, at first iteration $\mathbf{F}=\mathbf{F}_{\textnormal{HF}}$, and $\mathbf{\Sigma}(\omega)=\mathbf{0}$.

\subitem\textbf{2.a} Find $\mu$ such that $\mathbf{P}(\mathbf{G})$ has good particle number.
\subitem\textbf{2.b} Rebuild $\mathbf{F}(\mathbf{P})$ according to Eq.~\ref{eq:FockMatrix}, and update $\mathbf{G}(\omega)$. Repeat \textbf{2.a}-\textbf{2.b} until  $\mathbf{G}(\omega)$, $\mathbf{F}$, and $\mu$ are self-consistent. FFT $\mathbf{G}(\omega)\rightarrow\mathbf{G}(\tau).$
\item[\textbf{3.}] Build $\mathbf{\Sigma}(\tau)$ according to Eq.~\ref{eq:Sigma}. FFT $\mathbf{\Sigma}(\tau)\rightarrow\mathbf{\Sigma}(\omega)$. Go to step 2 and repeat until convergence.
\end{itemize}

Now we comment on the convergence properties of the GF2 algorithm. 
%Inspecting Fig.~\ref{fig:loops}, the GF2 algorithm can be viewed as two stages: 
%\begin{itemize}
%\item[\textbf{i.}] an ``inner loop'', where for a given fixed $\mathbf{\Sigma}$, the chemical-potenial $\mu$, $\mathbf{G}$, and $\mathbf{F}$ are iterated to self-consistency.
%\item[\textbf{ii.}] an ``outer loop'', where for a given $\mathbf{G}$ a new self-energy $\mathbf{\Sigma}$ is built.
%\end{itemize}
%
%\noindent
 In cases where the system is essentially single-reference, iterating to convergence is straightforward and one may allow loose self-consistency requirements on $\mathbf{G}(\omega)$, $\mathbf{F}$, and $\mu$ during the inner-loop in order to accelerate the calculation. In multi-reference cases however we find that applying tighter convergence criteria  on $\mathbf{G}(\omega)$, $\mathbf{F}$, and $\mu$ while damping $\mathbf{P}$ can stabilize convergence of the entire GF2 calculation.
For the outer loop we find it useful to apply damping to $\mathbf{\Sigma}(\tau)$, with the amount of damping depending on how multireference the system is.

%\clearpage

\subsection{Plots of Natural occupation numbers}

\begin{figure}[h]
\centerline{\includegraphics[width= 8.5cm]{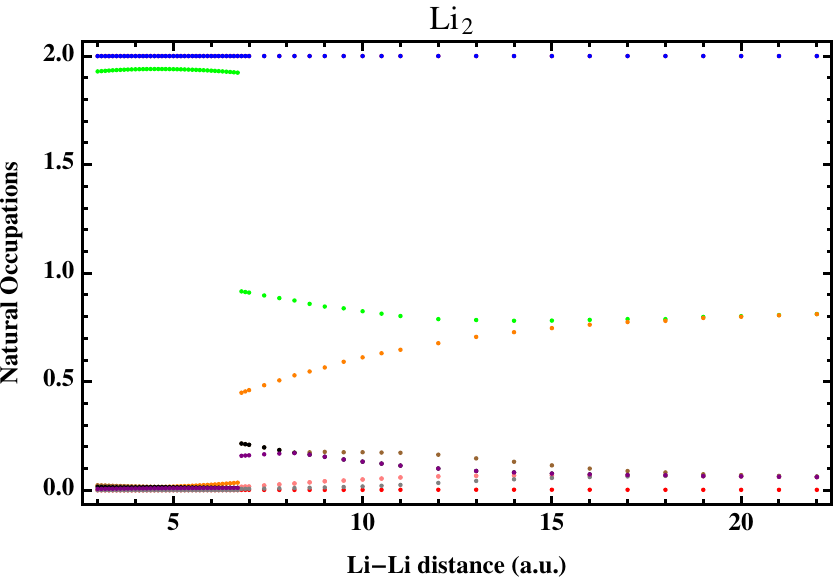}}
\caption{Natural occupation numbers of Li$_{2}$ as calculated by GF2 with respect to Li-Li bond distance.}
\label{fig:svwn5SC}
\end{figure}

\begin{figure}
\centerline{\includegraphics[width= 8.5cm]{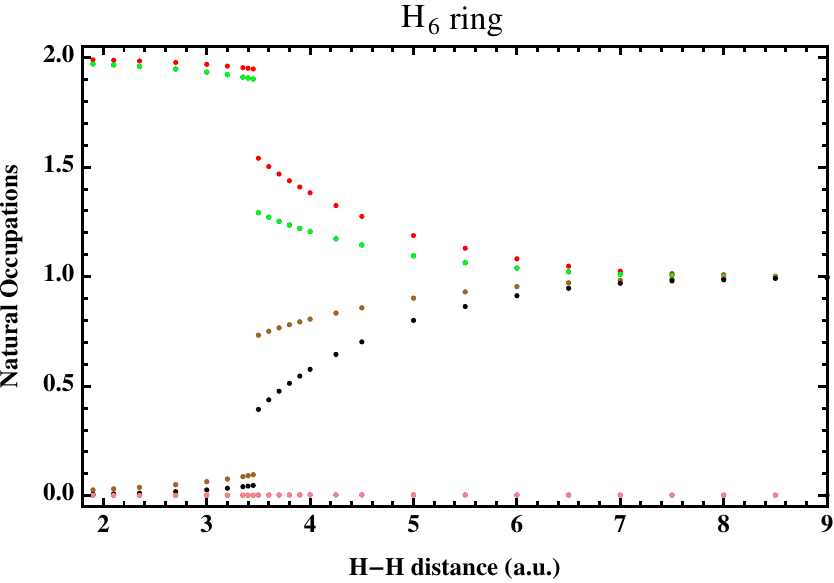}}
\caption{Natural occupation numbers of the H$_{6}$ ring as calculated by GF2 with respect to H-H distance.}
\label{fig:svwn5NSC}
\end{figure}

\begin{figure}[htp]
\centerline{\includegraphics[width= 8.5cm]{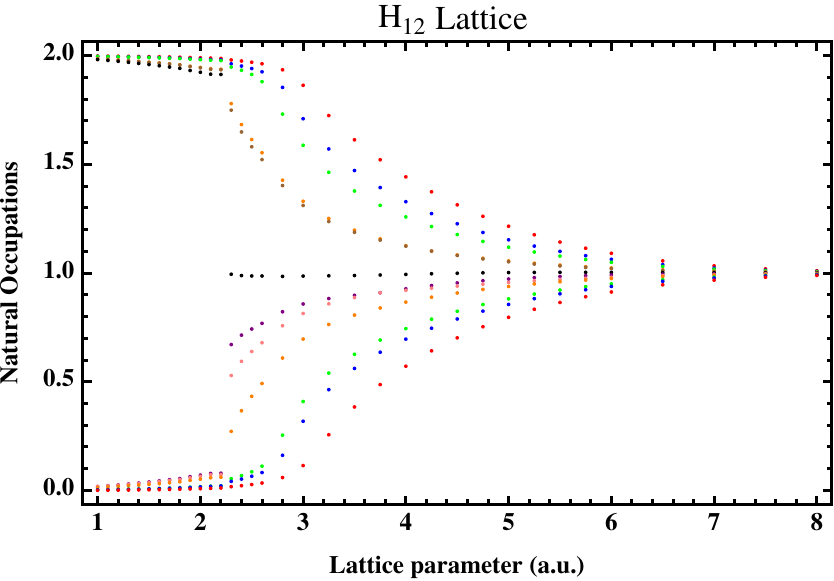}}
\caption{Natural occupation numbers of the 4$\times$3 H$_{12}$ lattice  as calculated by GF2 with respect to lattice parameter.}
\label{fig:pbevwn5SC}
\end{figure}
\newpage

%\bibliography{Bibliography}

\begin{thebibliography}{80}
\bibitem [{foo()}]{footnote}%
  \BibitemOpen
  \href@noop {} {}\bibinfo {note} {{We based this conclusion on the work of
  Scuseria where inclusion of mosaic diagrams prevented divergence of
  perturbative scheme for electron gas. Since mosaic diagrams are included in
  GF2 resummation, we conclude that such method must work for periodic metallic
  systems.}}\BibitemShut {Stop}%
\end{thebibliography}
%

\end{document}